\documentstyle[aps,epsfig,twocolumn]{revtex}

\title{Noise of Entangled Electrons: Bunching and Antibunching}

\author{Guido Burkard, Daniel Loss and Eugene V. Sukhorukov}
\address{
Department of Physics and Astronomy, University of Basel,\\
Klingelbergstrasse 82, CH-4056 Basel, Switzerland}

\begin{document}

\draft
\twocolumn[\hsize\textwidth\columnwidth\hsize\csname
@twocolumnfalse\endcsname

\maketitle

\begin{abstract}
Addressing the feasibility of quantum communication with entangled electrons
in an interacting many-body environment, we propose an interference experiment
using a scattering set-up with an entangler and a beam splitter.
It is shown that, due to electron-electron interaction, the fidelity of the entangled
singlet and triplet states is reduced by $z_F^2$ in a conductor described by Fermi
liquid theory. We calculate the quasiparticle weight factor $z_F$ for a
two-dimensional electron system.
The current noise for electronic singlet states turns out to be enhanced
(bunching behavior), while it is reduced for triplet states (antibunching).
Within standard scattering theory, we find that the Fano factor
(noise-to-current ratio) for singlets is twice as large as for
independent classical particles and is reduced to zero for triplets.
\end{abstract}


\vskip1pc]
\narrowtext

The availability of pairwise entangled qubits---Einstein-Podolsky-Rosen
(EPR) pairs\cite{Einstein}---is a necessary prerequisite for secure quantum
communication\cite{Bennett84}, dense coding\cite{Bennett92}, and
quantum teleportation\cite{Bennett93}. The prime example of an EPR pair
considered here is the singlet state formed by two electron spins, its
main feature being its non-locality: If the two entangled electrons are
separated from each other in space, then (space-like separated)
measurements of their spins are still strongly correlated, leading to
a violation of Bell's inequalities\cite{Bell}.
Experiments with photons have tested Bell's inequalities\cite{Aspect},
dense coding\cite{Mattle}, and quantum teleportation\cite{Zeilinger}.
To date, none of these phenomena have been seen for {\it massive}
particles such as electrons, let alone in a solid-state environment.
This is so because it is difficult to first produce and to second
detect entanglement of electrons in a controlled way.
On the other hand, recent experiments have demonstrated very long spin
decoherence times for electrons in semiconductors \cite{Kikkawa97}.
It is thus of considerable interest to see if it 
is possible to use mobile electrons in a many-particle system,
prepared in a definite (entangled) spin state, for the purpose of
quantum communication.

As to the production of entangled electrons, we have previously
described in detail how two electron spins can be deterministically
entangled by weakly coupling two nearby quantum dots, each of which
contains one single (excess) electron\cite{LD98,BLD99}.
As recently pointed out such a spin coupling can also be achieved 
on a long distance scale by using a cavity-QED scheme\cite{Imamoglu},
or with electrons which are trapped by surface acoustic waves on a
semiconductor surface\cite{Barnes}.

In this paper, we describe a method for \textit{detecting} pairwise
entanglement between electrons in two mesoscopic wires, which relies on
the measurement of the current noise in one of the wires.
For this purpose, we also study the propagation of entangled electrons
interacting with all other electrons in those wires (see further below).
Our main result is that the singlet EPR pair gives rise to an enhancement
of the noise power (``bunching'' behavior), whereas the triplet EPR pair
leads to a suppression of noise (``antibunching''). The underlying
physics responsible for this phenomenon is well known from the scattering
theory of two identical particles in vacuum\cite{Feynman,Ballentine}: in the
center-of-mass frame the differential scattering cross-section $\sigma$
can be expressed in terms of the scattering amplitude $f(\theta)$  and
scattering angle $\theta$ as
$\sigma(\theta)=|f(\theta) \pm f(\pi-\theta)|^2
=|f(\theta)|^2+|f(\pi-\theta)|^2 \pm 2 Re f^*(\theta)f(\pi-\theta)$.
The first two terms in the second equation are the ``classical''
contributions which would be obtained if the particles were
distinguishable, while the third (exchange) term results from their
indistinguishability which gives rise to genuine
two-particle interference effects.
\begin{figure}
\centerline{\psfig{file=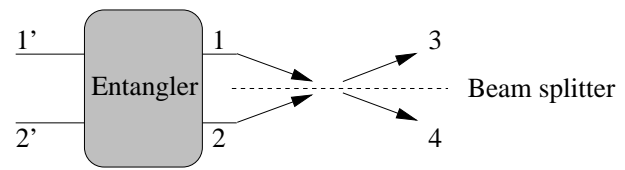,width=8cm}}
\vspace{3mm}
\caption{\label{fig1}
The setup for measuring the noise of entangled states.
Uncorrelated electrons are fed into the entangler (see text)
through the Fermi leads $1'$, $2'$ and are transformed into
pairs of electrons in the entangled singlet (triplet) state
$|\mp\rangle$, which are injected into leads $1$, $2$
(one electron of undetermined spin state into each lead).
The entanglement of the, say, spin singlet can then be detected
in an interference experiment using a beam splitter (with
no backscattering): Since the orbital wave function of
the singlet is symmetric, the electrons leave the scattering
region preferably in the same lead ($3$ or $4$). This correlation
(``bunching'') is revealed by an enhancement of the noise by a factor
of $2$ in the outgoing leads.}
\end{figure}
Here the plus (minus) sign applies to spin-1/2 particles in the
singlet (triplet) state, described by a (anti)symmetric orbital wave
function. The very same two-particle interference mechanism which
is responsible for the enhancement (reduction) of the scattering cross
section $\sigma(\theta)$ near $\theta=\pi/2$ also leads to an increase
(decrease) of the correlations of the particle number in the output
arms of a beam splitter\cite{Loudon}.

We turn now to the question of how to detect entanglement of electrons
in a solid-state environment. 
For this we propose a
non-equilibrium transport measurement 
using the set-up  shown in
Fig.~\ref{fig1}. Here, the entangler is assumed to be a 
device by which entangled
 states of two electrons can be generated, a specific realization
being above-mentioned double-dot system\cite{LD98,BLD99}.
The presence of a beam splitter ensures that the electrons leaving
the entangler have a finite amplitude to be interchanged
(without mutual interaction). 
Below we will show
that in the absence of spin scattering the noise measured in the
outgoing leads 3 and 4 will exhibit bunching behavior for pairs of electrons
with a symmetric orbital wave function\cite{noise}, i.e. for spin
singlets, while antibunching behavior is found in the case of the
spin triplets, due to their antisymmetric orbital wave function.
The latter situation is the one considered so far for electrons in the
normal state both in theory\cite{Buettiker1,Martin} and in
recent experiments\cite{Stanford,MartinSC}. These experiments \cite{Stanford}
have been performed in semiconducting nanostructures with geometries that are
closely related to the set-up proposed in Fig.~\ref{fig1} but without entangler.
Note that since the (maximally entangled) singlet is the
only state leading to  bunching behavior, the latter effect can be
viewed as a unique signature for the entanglement of the injected
electrons. 
To establish these results we first need to  assess the effect of
interactions in the leads.
Thus we proceed in two steps: First, we show that the entanglement
of electrons injected into Fermi leads is only partially degraded by
electron-electron interactions. This allows us then to  
use, in a second step, the standard
scattering matrix approach\cite{Buettiker1}---which we extend to a
situation with entanglement---in terms of
(non-interacting) Fermi liquid quasiparticles.

{\it Entangled electrons in a Fermi liquid.}
Electrons are injected from the entangler
(say, a pair of coupled quantum dots) into the leads,
e.g.\ by (adiabatically) lowering the gate barriers between dot
and lead, in the spin triplet (upper sign) or singlet (lower sign)
state,
\begin{equation}
|\psi_{{\bf n}{\bf n'}}^{t/s} \rangle={1\over \sqrt{2}}
(a_{{\bf n}\uparrow}^\dagger a_{{\bf n'}\downarrow}^\dagger \pm
a_{{\bf n}\downarrow}^\dagger\, a_{{\bf n'}\uparrow}^\dagger\,)\, |\psi_0\rangle,
\label{state}
\end{equation}
with ${\bf n}=({\bf q},l)$, where ${\bf q}$ is the momentum of an electron,
and $l$ is the lead number.
Here, $\psi_0$ denotes the 
filled Fermi sea, the electronic ground state in the leads, and we
have used the fermionic creation ($a^\dagger_{{\bf n}\sigma}$) and annihilation
($a_{{\bf n}\sigma}$) operators, where $\sigma$ denotes spin in the
$\sigma_z$-basis.
Next, we introduce the transition
amplitude $G^{t/s}({\bf 1}{\bf 2},{\bf 3}{\bf 4};t)
=\langle\psi^{t/s}_{{\bf 1}{\bf 2}}, t|\psi^{t/s}_{{\bf 3}{\bf
4}}\rangle$ and 
define the \textit{fidelity} as the modulus of 
$G^{t/s}$ between the same initial and final states,
$|G^{t/s}({\bf 1}{\bf 2},{\bf 1}{\bf 2};t)|
=|G^{t/s}({\bf 2}{\bf 1},{\bf 1}{\bf 2};t)|$.
The fidelity is a measure of how much of the initial triplet (singlet)
remains in the final state after propagating for time $t>0$ in a Fermi
sea (metallic lead) of interacting electrons.
We emphasize that after injection, the two
electrons of interest are no longer distinguishable from the
electrons of the leads, and consequently the two electrons taken out of
the leads will, in general, not be the same as the ones injected.
Introducing the notations $n=({\bf n},\sigma)$,
and $\bar{n}=({\bf n},-\sigma)$, we write
\begin{equation}
G^{t/s}({\bf 1}{\bf 2},{\bf 3}{\bf 4};t)
=-{1\over 2}\sum_{\sigma} \!\left[ G( 1 \bar{2}, 3 \bar{4};t) 
\pm G( 1\bar{2}, \bar{3} 4;t)\right] ,
\label{generaloverlap}
\end{equation}
where we introduced the standard 2-particle Green's function $G( 1 2, 3 4;t)
=\langle \psi_0| T a_1(t) a_2(t)
a_3^\dagger a_4^\dagger|\psi_0\rangle$, with $T$ the time ordering operator.
We assume a time- and spin-independent
Hamiltonian, $H=H_0+\sum_{i<j} V_{ij}$, where $H_0$ describes the free
motion of the $N$ electrons, and $V_{ij}$ is the bare Coulomb
interaction between electrons $i$ and $j$.

The non-trivial many-body problem of finding an explicit value
for $G(12,34;t)$ is simplified because we can assume that the leads
$1$ and $2$ are sufficiently separated, so that the mutual Coulomb
interaction can be safely neglected.
This implies that the 2-particle vertex part vanishes and we
obtain $G(12,34;t)=G(13,t)G(24,t)-G(14,t)G(23,t)$, 
i.e. the Hartree-Fock approximation is exact and the problem is reduced
to the evaluation of single-particle Green's functions 
$G({\bf n},t)
=- i\langle \psi_0| T a_{{\bf n}}(t)a_{{\bf n}}^\dagger|\psi_0\rangle
\equiv G_l({\bf q},t)$, 
pertaining to lead $l=1,2$ (the leads are still interacting
many-body systems though). Inserting this into Eq.~(\ref{generaloverlap})
we arrive at  the result
$G^{t/s}({\bf 1}{\bf 2},{\bf 3}{\bf 4};t)= -
G({\bf 1},t)\, G({\bf 2},t)[ \delta_{{\bf 13}}\delta_{{\bf 24}}
\mp \delta_{{\bf 14}} \delta_{{\bf 23}}]$,
where the upper (lower) sign refers to the spin triplet (singlet).
For the special case $t=0$, and no interactions, we have $G({\bf n},t)=-i$,
and thus $G^{t/s}$ reduces to $ \delta_{{\bf 13}}\delta_{{\bf 24}}
\mp \delta_{{\bf 14}} \delta_{{\bf 23}}$,
and the fidelity is $1$.
In general, we have to evaluate the (time-ordered) single-particle
Green's functions $G_{1,2}$ close to the Fermi surface and obtain the standard
result\cite{Mahan}
$G_{1,2}({\bf q},t) \approx -i z_q
\Theta(\epsilon_q -\epsilon_F) e^{-i \epsilon_q t- \Gamma_qt}$,
which is valid for $0\leq t\lesssim 1/\Gamma_q$, where $1/\Gamma_q$ is the
quasiparticle lifetime, $\epsilon_q=q^2/2m$ the quasiparticle energy
(of the added electron), and $\epsilon_F$ the Fermi energy.
For a two-dimensional electron system (2DES),
as e.g.\ in GaAs heterostructures,
$\Gamma_q \propto (\epsilon_q -\epsilon_F)^2
\log(\epsilon_q -\epsilon_F)$ \cite{Quinn} within the random phase
approximation (RPA), which accounts for screening and which is
obtained by summing all polarization diagrams\cite{Mahan}.  Thus, the
lifetime becomes infinite when the energy of the added electron
approaches $\epsilon_F$ (with Fermi momentum $k_F$).
The most important quantity in the present context is the
renormalization factor or quasiparticle weight, $z_F=z_{k_F}$,
evaluated at the Fermi surface, defined by
$z_F=[1-\partial/\partial \omega \,
{\rm Re}\Sigma( k_F,\omega=0)]^{-1}$,
where $\Sigma(  q,\omega)$ is the irreducible self-energy.
The quasiparticle weight, $0\leq z_{q}\leq 1$, describes
the weight of the bare electron in the quasiparticle state ${\bf q}$.
For momenta ${\bf q}$ close to the Fermi surface and for identical leads
($G_1=G_2$) we find
\begin{equation}
|G^{t/s}({\bf 1}{\bf 2},{\bf 3}{\bf 4};t)|
=\, z_F^2 \, |\, \delta_{{\bf 13}}\delta_{{\bf 24}}
\mp \delta_{{\bf 14}} \delta_{{\bf 23}}|
\end{equation}
for all times satisfying $0< t\lesssim 1/\Gamma_q$.
Thus we find that the fidelity for
singlet and triplet states in the presence of a Fermi sea and
Coulomb interaction is given by $z_F^2$. Since this is the
sought-for measure of the reduction of the spin correlation,
it is interesting to evaluate $z_F$ explicitly for a 2DES.
Evaluating $\Sigma$
within RPA (and imaginary time), we obtain
${\Sigma} ({\bar k}) = -(1/\Omega\beta)\sum_{\bar q} G^0({\bar
k}+{\bar q})v_{ q}\varepsilon({\bar q})$, where $\beta=1/k_BT$ is
the inverse temperature, 
$\Omega$ the volume 
and ${\bar q}=(q_n,{\bf q})$, with $q_n$  the Matsubara
frequencies. The unperturbed Green's function
is given by $G^0({\bar q})=(iq_n-\xi_{\bf q})^{-1}$, where
$\xi_{\bf q}=\varepsilon_{\bf q}-\varepsilon_F$, and the 
Coulomb interaction in
two dimensions is $v_{ q}=2\pi e^2/{q}$.
The dielectric function can then be expressed as
$\varepsilon = \varepsilon_0 - v_{ q} P^{(1)}({\bar q})$, using
the polarization propagator in leading order,
$P^{(1)}({\bar q}) = -\Omega^{-1}\sum_{{\bf p},\sigma}[n_F(\xi_{\bf
p})-n_F(\xi_{\bf p+q})]/[\xi_{\bf p}-\xi_{\bf p+q}+iq_n]$, where
$n_F(\xi_{\bf p})=(e^{\beta\xi_{\bf p}}+1)^{-1}$.
In two dimensions, we find
$P^{(1)}({\bar q}) = (2mk_F/\pi q) {\rm Re} (\sqrt{u^2-1}-u)$,
with $u={ q}/2k_F+im\omega/{ q}k_F$, and where the branch cut of
$\sqrt{u^2-1}$ is on $[-1,1]$.
After careful analytic continuation \cite{Mahan} and some lengthy
calculation, we finally obtain
\begin{equation}
z_F=1-r_s \left({1\over 2} +{1\over \pi}\right),
\label{qpweight}
\end{equation}
in leading order of the interaction parameter $r_s=1/k_F a_B$, where
$a_B=\epsilon_0\hbar^2/me^2$ is the Bohr radius.  In particular, in a
GaAs 2DES we have $a_B=10.3$ nm, and $r_s=0.614$, and thus we obtain
$z_F=0.665$ \cite{footnotenumerical,3d}.

We see that the fidelity (spin correlation) is reduced by a factor
of $z_F^{-2}\approx 2$ (from its maximum value $1$) as soon as we inject
the two electrons (entangled or not) into separate leads consisting of {\it
interacting} Fermi liquids in their ground state.
Apart from this reduction, however, the entanglement is not affected by
interacting electrons in the filled Fermi sea.
This result allows us now to study the noise of entangled electrons
using the standard scattering theory for quasiparticles
in a Fermi liquid\cite{footnote2}.

{\it Noise of entangled electrons.}
We now investigate the noise correlations for scattering with the
entangled incident state
$|\pm\rangle \equiv |\psi_{{\bf 1}{\bf 2}}^{t/s}\rangle$,
where we set ${\bf n}=(\varepsilon_n,n)$,
now using the electron energies $\varepsilon_n$ instead of the
momentum as the orbital quantum number in Eq.~(\ref{state}) and
where the operator
$a^\dagger_{\alpha \sigma}(\varepsilon)$ creates an incoming
electron in lead $\alpha$ with spin $\sigma$ and energy
$\varepsilon$.
(Another interesting spin effect is noise induced by spin
transport\cite{spinnoise}.)
First, we generalize the theory for the current correlations in a
multiterminal conductor as given in Ref.~\onlinecite{Buettiker1}
to the case of entangled scattering states, with the important consequence
that Wick's theorem cannot be applied directly.
We start by writing the operator for the current carried by electrons
in lead $\alpha$ of a multiterminal conductor as
\begin{eqnarray}
I_{\alpha}(t) =
\frac{e}{h\nu} \sum_{\varepsilon \varepsilon ' \sigma} \left[
a_{\alpha \sigma}^\dagger(\varepsilon)a_{\alpha \sigma}(\varepsilon ')
- b_{\alpha \sigma}^\dagger(\varepsilon)
b_{\alpha \sigma} (\varepsilon ')\right]&&
\nonumber\\
\times \exp\left[i(\varepsilon-\varepsilon ')t/\hbar\right], &&
\label{current_def}
\end{eqnarray}
where the operators
$b_{\alpha \sigma}(\varepsilon)$ for the outgoing electrons are related to the operators
$a_{\alpha \sigma}(\varepsilon)$ for the incident electrons via
$b_{\alpha\sigma}(\varepsilon)=\sum_{\beta} s_{\alpha\beta}a_{\beta \sigma}(\varepsilon)$, 
where
$s_{\alpha\beta}$ denotes the scattering matrix.
We assume that the scattering matrix is spin- and energy-independent.
Note that since we are dealing with discrete energy states here, we
normalize the operators $a_{\alpha \sigma}(\varepsilon)$ such that
$\{a_{\alpha \sigma}(\varepsilon),a_{\beta \sigma'}^\dagger(\varepsilon ')\}=
\delta_{\sigma\sigma'}\delta_{\alpha\beta}\delta_{\varepsilon\varepsilon '}/\nu$,
where the Kronecker symbol $\delta_{\varepsilon\varepsilon '}$ equals $1$ if 
$\varepsilon=\varepsilon '$ and $0$
otherwise.
Therefore we also have to include the factor $1/\nu$ in the definition of
the current, where $\nu$ stands for the density of states in the leads.
We assume that each lead consists of only a single quantum
channel;
the generalization to leads with several channels is straightforward but is
not needed here.
Using the scattering matrix, we can write
Eq.~(\ref{current_def}) as
\begin{eqnarray}
  I_{\alpha}(t) = \frac{e}{h\nu}\sum_{\varepsilon \varepsilon ' \sigma} 
  \sum_{\beta\gamma}
 a_{\beta \sigma}^\dagger (\varepsilon) A_{\beta\gamma}^\alpha
a_{\gamma \sigma}(\varepsilon ') e^{i(\varepsilon-\varepsilon ')t/\hbar} , 
&&\label{current}\\
A_{\beta\gamma}^{\alpha} =
\delta_{\alpha\beta}\delta_{\alpha\gamma}
-s_{\alpha\beta}^{*} s_{\alpha\gamma}. 
&&\label{A2}
\end{eqnarray}
The spectral density of the current fluctuations (noise)
$\delta I_{\alpha}=I_{\alpha}-\langle I_{\alpha}\rangle$
between the leads $\alpha$ and $\beta$ is defined as
\begin{equation}
  \label{cross1}
  S_{\alpha\beta}({\omega})
  = \lim_{T\rightarrow\infty}
  \frac{h\nu}{T}\int_0^T\!\!\!dt\,\,e^{i\omega t}
  \,{\rm Re}\,
  \langle\pm|\delta I_{\alpha}(t)\delta I_{\beta}(0)|\pm\rangle.
\end{equation}
We evaluate now the correlations Eq.~(\ref{cross1}) for zero frequency.
Using the fact that the unpolarized currents are invariant when
all spins are reversed, the expectation value
$\langle\pm|\delta I_{\alpha} \delta I_{\beta}|\pm\rangle$
can be expressed as the sum of a direct and an exchange term,
$\langle\pm|\delta I_{\alpha} \delta I_{\beta}|\pm\rangle
  =  \langle\uparrow\downarrow\!|\delta I_{\alpha} \delta I_{\beta}|\!
  \uparrow\downarrow\rangle
  \pm\langle\uparrow\downarrow\!|\delta I_{\alpha} \delta I_{\beta}|\!
  \downarrow\uparrow\rangle$,
where the upper (lower) sign of the exchange term refers to triplet
(singlet).
Evaluating these expressions further, we arrive at the following result
for the zero-frequency ($\omega=0$) correlation between the leads $\alpha$
and $\beta$,
\begin{equation}
  S_{\alpha\beta}\!
   = \!\frac{e^2}{h\nu}\Big[\sum_{\gamma\delta}\!{}^{'}
    A_{\gamma\delta}^{\alpha}A_{\delta\gamma}^{\beta}
   \mp \delta_{\varepsilon_1,\varepsilon_2}
    \big(A_{12}^{\alpha}A_{21}^{\beta}\! +\!A_{21}^{\alpha}A_{12}^{\beta}
\big)\Big],\label{cross5}
\end{equation}
where $\sum_{\gamma\delta}^{\prime}$ denotes the sum over $\gamma=1,2$ and
all $\delta\neq\gamma$, and where again the upper (lower) sign refers to 
triplets (singlets).
The autocorrelations $S_{\alpha\alpha}$ determine the noise in lead
$\alpha$ (note that $A_{\gamma\delta}^{\alpha}A_{\delta\gamma}^{\alpha}
=|A_{\gamma\delta}^{\alpha}|^2$).

We apply our result Eq.~(\ref{cross5}) to the set-up shown in
Fig.~\ref{fig1} involving four leads, described by the single-particle
scattering matrix elements, $s_{31}=s_{42}=r$, and $s_{41}=s_{32}=t$,
where $r$ and $t$ denote the reflection and transmission amplitudes
at the beam splitter, respectively.
We assume that there is no backscattering,
$s_{12}=s_{34}=s_{\alpha\alpha}=0$.
The unitarity of the s-matrix implies $|r|^2+|t|^2=1$,
and ${\rm Re}[r^*t]=0$.
Using Eqs.~(\ref{A2}) and (\ref{cross5}), we obtain the final
result for the noise correlations for the
incident state $|\pm\rangle$\cite{footnote1},
\begin{equation}
  \label{noise}
S_{33}=S_{44}=-S_{34}=2\frac{e^2}{h\nu}T\left(1-T\right)
  \left(1\mp \delta_{\varepsilon_1\varepsilon_2}\right),
\end{equation}
where $T=|t|^2$ is the probability for transmission through the
beam splitter.
The calculation for the remaining two triplet states
$|\!\uparrow\uparrow\rangle$ and $|\!\downarrow\downarrow\rangle$
yields the same result Eq.~(\ref{noise}) (upper sign).
For the average current in lead $\alpha$ we obtain
$\left|\langle I_\alpha\rangle\right| = e/h\nu$,
with no difference between singlets and triplets.
Then, the Fano factor
$F = S_{\alpha\alpha} /\left|\langle I_\alpha\rangle\right|$
takes the following form
\begin{equation}
  \label{fano}
  F =  2eT(1-T)\left(1\mp \delta_{\varepsilon_1\varepsilon_2}\right),
\end{equation}
and correspondingly for the cross correlations.
Eq.~(\ref{fano}) is one of the main results of this work:
it implies that if two electrons with the same energies,
$\varepsilon_1=\varepsilon_2$, in the singlet
state $|-\rangle$ are injected into the leads $1$ and $2$,
then the zero frequency noise is {\it enhanced} by a factor of two,
$F=4eT(1-T)$, compared to the shot noise of uncorrelated particles
\cite{Buettiker1,noisesuppression},
$F=2eT(1-T)$. This enhancement of noise is
due to {\it bunching} of electrons in the outgoing leads, caused by the
symmetric orbital wavefunction of the spin singlet $|-\rangle$.
On the other hand, the triplet states $|+\rangle$
exhibit {\it antibunching}, i.e. a complete suppression of the noise,
$S_{\alpha\alpha}=0$.
The noise enhancement for the singlet $|-\rangle$ is a
unique signature for entanglement (there exists no unentangled state with
the same symmetry), therefore entanglement can be observed by
measuring the noise power of a mesoscopic conductor
as shown in Fig.~\ref{fig1}.
The triplets $|+\rangle$, $|\!\uparrow\uparrow\rangle$ and
$|\!\downarrow\downarrow\rangle$ can be distinguished from each
other by a measurement of the spins of the outgoing electrons,
e.g. by inserting spin-selective tunneling devices\cite{Prinz} into
leads $3$ and $4$.

In conclusion, we have demonstrated that
entangled electrons (EPR pairs) can be transported in mesoscopic wires,
and we have quantified the reduction of entanglement during this process.
The current fluctuations in a beam-splitter set-up turn
out to be a suitable experimental probe for detecting (entangled)
spin states of electrons via their charge.

\acknowledgements
We would like to thank M. B\"uttiker and D. DiVincenzo for useful
discussions.
This work has been supported by the Swiss National Science Foundation.


\begin{references}
\bibitem{footnotenumerical} 
After a more accurate numerical evaluation of the
exact RPA self-energy, we find for GaAs $z_F=0.691155$.
\bibitem{footnote2} As in the unentangled case
\cite{Buettiker1,Martin,noisesuppression}, we assume that the noise
correlations that we find using the scattering-matrix approach are 
at most quantitatively but not qualitatively altered by
Coulomb interactions.
\bibitem{spinnoise}
Noise induced by spin currents can be observed 
in a two-terminal conductor attached to Fermi leads with spin-dependent
bias $\Delta\mu_{\sigma}$.
We have $\left<I_{\sigma}\right>={e\over h}T\Delta
\mu_{\sigma}$, and from Eq.\ (\ref{current_def}) we obtain the noise power
$S={{e^2}\over h}T(1-T)\left(\left|\Delta\mu_{\uparrow}\right
|+\left|\Delta\mu_{\downarrow}\right|\right).$
In particular, when $\Delta\mu_{\uparrow}=\Delta\mu_{\downarrow}$
we obtain the usual result\cite{noisesuppression} $S=e(1-T)\left|I_c\right|$
for the shot noise induced by the charge current
$I_c\equiv\left<I_{\uparrow}\right>+\left<I_{\downarrow}\right>$.
On the other hand, for
$\Delta\mu_{\uparrow}=-\Delta\mu_{\downarrow}$, i.e.\ when there
is no charge current through the
conductor, $I_c=0$, but still there is a non-vanishing spin current
$I_s\equiv\left<I_{\uparrow}\right>-\left<I_{\downarrow}\right>$, one can
observe the current noise $S=e(1-T)\left|I_s\right|$
induced by spin transport only.
\bibitem{footnote1}
For finite frequencies, we obtain the noise power
$S_{\alpha\alpha}(\omega) = S_{\alpha\alpha}^{\rm FS}(\omega)
 +   (e^2/h\nu)[(1-\delta_{\omega,0})
 +   T(1-T)(2\delta_{\omega,0}
 \mp \delta_{\omega, \varepsilon_1-\varepsilon_2}
 \mp \delta_{\omega, \varepsilon_2-\varepsilon_1}
)]$.
The noise contribution $S_{\alpha\alpha}^{\rm FS}$ due to the Fermi sea
is independent of the spin state of the injected pair.
\end{references}
\end{document}